# MECHANICAL ENERGY FLUXES ASSOCIATED WITH SATURATED CORONAL HEATING IN M DWARFS: COMPARISON WITH PREDICTIONS OF A TURBULENT DYNAMO

Short title: Mechanical fluxes and coronal heating in low-mass stars

D. J. Mullan

Bartol Research Institute, Dept of Physics and Astronomy, University of Delaware, Newark DE 19716: mullan@udel.edu

and

J. MacDonald

Dept of Physics and Astronomy, University of Delaware, Newark DE 19716

## ABSTRACT

Empirically, the X-ray luminosity $L_X$ from M dwarfs has been found to have an upper limit of about 0.2% of the bolometric flux $L_{bol}$. In the limit where magnetic fields in M dwarfs are generated in equipartition with convective motions, we use stellar models to calculate the energy flux of Alfven waves $F_A$ as a function of depth in the sub-surface convection zone. Since Alfven waves have the optimal opportunity for wave modes to reach the corona, we suggest that $F_A$ sets an upper limit on the mechanical flux $F_{mech}$ which causes coronal heating. This suggestion accounts quantitatively for the "saturated" values of $L_X/L_{bol}$ which have been reported empirically for M dwarfs.

*Key words*: stars: activity - stars: coronae - stars: low-mass - stars: magnetic field

1. INTRODUCTION

Many late-type stars emit radiant energy of a non-thermal nature from chromospheres and coronae. These non-thermal processes can be quantified by ratios such as $L_{HK}/L_{bol}$ (from the chromospheric H and K lines) and $L_X/L_{bol}$ (using the coronal X-rays). These ratios can be considered as measures of the efficiencies with which the total output power of the star is converted into chromospheric heating and coronal heating, respectively. Different stars emit at different intensities, with ranges spanning 4 or more orders of magnitude in $L_X/L_{bol}$. The data indicate that a well-defined *upper* limit exists, above which no cool dwarf (so far) has been found to emit coronal X-rays. The upper boundary of the range of emission in chromospheric and/or coronal emissions is referred to as the "saturated level". Vilhu & Walter (1987: VW) provide data for coronal emission from several dozen stars ranging from spectral type A to M5 and later. The M dwarfs plotted by VW are taken from a compilation by Bookbinder (1985) of the most active M dwarfs within 25 parsecs.

The upper limit on $L_X/L_{bol}$ for the most active M dwarfs reported by VW is about $2 \times 10^{-3}$, indicating that of the entire energy flux emerging from one such star, some 0.2% is converted into coronal heating. To convert to actual fluxes of X-rays, we note that the photospheric fluxes from M dwarfs with effective temperatures in the range 4000-3300 K (corresponding to spectral sub-types M1-M5 according to five different calibrations of M dwarfs illustrated by Muzic et al. 2014) are in the range $(1.5-0.7) \times 10^{10}$ erg cm$^{-2}$ s$^{-1}$. A fraction 0.2% corresponds to upper limits on the coronal heating fluxes in the range $(1.4-3) \times 10^7$ erg cm$^{-2}$ s$^{-1}$. We note that although M dwarfs have the largest values of the ratio $L_X/L_{bol}$ in the VW dataset, the largest absolute values of $L_X$ itself (reaching almost as large as $10^8$ erg cm$^{-2}$ s$^{-1}$) are found among early G stars.

The results of VW indicate that there is a clear distinction between stars in which the corona is at the "saturated level" and stars where the coronal heating is "unsaturated". On the one hand, in the "unsaturated" stars, the data show that the $L_X/L_{bol}$ value rises steeply as the stellar rotation period $P$ becomes shorter (see Fig. 10 in VW): in a log-log plot of $L_X/L_{bol}$ versus Rossby number Ro (= $\tau_c /P$, where $\tau_c$ is the convective turnover time), the slope $\beta$ has a (positive) value of 3 or more. Thus, the rotation-activity correlation (RAC), i.e. $L_X/L_{bol} \sim P^a$, is expected to have a steep negative slope $a$ for the unsaturated stars. However, even if $\beta$ is known for a sample of stars with a given spectral type, the actual numerical value of $a$ for that sample depends on having access to a method which allows us to remove the variations of $\tau_c$ with spectral type. On the other hand, for the "saturated" stars, VW find that the value of $L_X/L_{bol}$ is essentially independent of Ro. That is, the RAC for saturated stars has a slope $a$ which is essentially zero.

Other examples of "saturation" in X-ray fluxes generated by the stars with fastest rotations have been provided by Micela et al. (1985), James et al. (2000), Pizzolato et al. (2003), and Wright et al. (2011). The sample of X-ray emitting stars presented by Wright et al. (2011: their Fig. 2) contains 824 stars, an order of magnitude larger than the VW sample. But the Wright et al. sample confirms the VW result that the RAC has a clearly defined slope of zero among the saturated stars. The latter have an upper limit on $L_X/L_{bol}$ of (2-3) x $10^{-3}$, essentially identical to the upper limit reported by VW.

Among the *unsaturated* stars in the Wright et al. sample, a fit to the Rossby number indicates a well-defined (positive) slope of 2.7, comparable to the slope obtained by VW. Wright et al. also present a log-log plot of $L_X/L_{bol}$ versus *P*: in this case, there is more scatter in the data: an "eyeball" fit data suggests that a functional form $L_X/L_{bol} \sim P^a$ leads to *a* values which are at least as steep as -1.

The aim of the present paper is to consider a quantitative explanation for the existence of a mechanical flux of up to (1.4-3) x $10^7$ erg $cm^{-2}$ $s^{-1}$ originating in M dwarfs. Moreover, the flux must be in a form which can cause deposition of heating energy at distances which are relatively far from the source. In particular, the flux must be in a form which allows the energy to propagate all the way up to altitudes which are relevant to the *corona*.

The main focus of the present paper is on M dwarfs with masses near to (or later than) the transition between partially convective stars and fully convective stars. The reason for this focus is that we have recently (Mullan et al. 2015: MHM) presented a calculation of dynamo action in stars where a specific physical structure exists inside the star: an *interface* between convective envelope and radiative core. In such stars, MHM quantify the possibility that dynamo action can be driven by rotational motion *at the interface*. In the present paper, we switch to consideration of a dynamo which relies for its operation only on the presence of convective motions. Such a dynamo can operate in any cool star where a convective envelope exists, *whether or not the star also contains an interface*. In its purest form, the dynamo we consider here operates in stars which are completely convective: such stars have no access to an interface dynamo, and therefore rely solely on the dynamo we consider here. In principle, it is possible that both types of dynamos may operate simultaneously in partially convective stars. Therefore, the results we present here are not meant to apply exclusively to fully convective stars. But here, in order to keep the discussion as simple as possible, we focus on a dynamo which has no need of an interface at all. In this sense, the present paper is meant to serve as a complement to MHM.

2. HEATING OF THE *CHROMOSPHERE* IN COOL DWARFS

The ultimate source(s) of chromospheric and coronal heating is (are) fluxes of mechanical energy which originate in a variety of modes in the convective zone. Whatever the mode of energy, the key point which we wish to emphasize here is that the energy must find a way to propagate up into the chromosphere or corona before dissipating.

In view of the existence of highly time-dependent convective motions in the sub-photospheric envelopes of cool dwarfs, it is inevitable that a certain flux of acoustic wave energy is generated by the time-dependent convection cells (e.g. Mullan 2009). Acoustic waves with periods which are shorter than the acoustic cut-off will propagate upwards from the convection zone into the upper photosphere. As regards the chromosphere, this region of the atmosphere is close enough to the photosphere (within a few scale heights) that mechanical energy emerging in the form of acoustic waves is suitable for the task of chromospheric heating. The quantitative reason for this claim is that acoustic waves can be modeled without ambiguity in a medium with specified temperature and density. Hydrodynamic modeling of this kind, using physically realistic boundary conditions based on the method of characteristics, and radiative cooling based on NLTE line cooling, shows that the waves steepen in such a way that flux which is created in the photosphere propagates upwards and dissipates over altitudes of no more than a few scale heights (e.g. Mullan & Cheng [1994], especially their Fig. 5). Acoustic heating of gas with chromospheric densities can reach temperatures of order (0.5-1) x $10^4$ K, sufficient to account for observed energy fluxes of CaII H and K lines. Although some remnants of acoustic waves may persist to high enough altitudes, thereby leading to temperatures which approach $10^6$ K (if the density is low enough), such high temperatures give rise to highly effective refraction of the acoustic waves. So serious is the refraction that the fluxes of acoustic waves which might ultimately be called upon to dissipate in the corona are essentially choked off before ever reaching the necessary altitudes.

It is not merely acoustic modes which give rise to chromospheric heating. Mechanical energy in other forms can also arise in magnetic regions, either in the form of MHD waves or as excess dissipation due to locally enhanced resistivity. These modes can give rise to localized excesses of chromospheric emission in "active regions". However, it is difficult to quantify these other excesses: there are many unknown free parameters which would need to be specified if we wished to quantify magnetic effects in chromospheric heating.

3. HEATING OF THE *CORONA* IN COOL DWARFS

As regards heating of the *corona*, stars may rely on waves, or on multiple magnetic reconnection events "nanoflares". Although nanoflares are of great current interest in the context of solar coronal heating (e.g. Bradshaw et al. 2012), quantitative modeling requires specification of many parameters, including the various dimensions of the current sheet, the strength of the local field, the rate of reconnection, the time-scale on which the nanoflares recur, and the radiative cooling time-scale of the ambient corona. In the case of the Sun, there may at some point in time be hope for narrowing down the ranges of these parameters. But in the case of other stars, reliable quantitative evaluation of the nanoflare process seems to be still a long way off.

In view of this, we prefer to focus on waves as the source of mechanical energy for coronal heating in cool dwarfs. But in order for waves to be relevant in any *coronal* process, the waves in question, assuming they have been generated in the convection zone, must inevitably satisfy a certain "survivability" condition: namely, they must be able to persist at significant distances from their source. To achieve this goal, the dissipation lengths of the waves must be relatively long so they can survive at distances that are far from their source. We have already mentioned (Section 2) that acoustic waves are not suitable in this regard: they dissipate too quickly. Therefore we rely on waves of a magnetic nature (i.e. MHD waves) to heat the corona. In coronal conditions, the physical processes which contribute to dissipation of MHD waves require us to consider two distinct limits.

First, at low enough densities, the conditions in the gas are collisionless. In such a case, linear theory suggests that fast and slow mode MHD modes (with their associated density fluctuations) are "very heavily damped through kinetic processes" (Matthaeus et al. 1999). As a result, the surviving fluctuations are expected to be preferentially in the Alfven mode. Evidence in favor of predominant Alfvenicity has been obtained in many regions of the (collisionless) solar wind (e.g. Belcher & Davis 1971; Smith et al. 2004).

Second, at higher densities, the coronal material behaves in a collisional manner, including processes such as Joule and frictional dissipation. In the corona, the protons are collisional between the surface and a finite altitude, perhaps 0.2-0.3 solar radii ($\approx 10^5$ km) above the surface (Endeve & Leer 2001). Can Alfven waves survive over distances of order $10^5$ km in a collisional corona? Osterbrock (1961) has shown that, in the presence of Joule dissipation and friction (both collisional processes), Alfven waves in active regions (where $B = 50$ G) have dissipation lengths of $10^5$ km (or more) provided they are launched from depths no lower than 250 km above the photosphere. Given the existence of overshooting above the top of the convection zone, it is reasonable to assume that some convective motions (which serve as mechanisms to "jostle" the field lines) can survive at such altitudes, within 1-2 scale heights of the formal top of the convection zone. Thus, Alfven waves *can* survive in the collisional corona. In contrast, slow mode MHD waves in the collisional portion of the corona (below 0.35 solar radii) are quickly damped (due to compressive viscosity and thermal conductivity) on length scales of no more than a few times $10^3$ km (Dwivedi & Srivastava 2008). Fast mode waves, with their compressional nature, are also expected to be damped on short length scales in the collisional corona.

These considerations indicate that Alfven waves have the ability to persist over great distances in both the collisional and collisionless regimes of a corona. In view of this result, we assume in the present paper that the waves which have the best chance of causing coronal heating in cool stars are Alfven waves. This is an essential feature of the results we present here.

4. PREVIOUS ATTEMPTS TO EXPLAIN SATURATED HEATING

Vilhu (1984) proposed that the saturation of chromospheric emission is due to the "total filling of the stellar disc with active regions". A quantitative examination of this proposal would require us to model not only how strong the fields are which emerge from dynamo operation, but also how much magnetic flux is generated. It is not clear that models of this kind, with the power to predict not only field strength but also the total amount of magnetic flux which is generated, are readily available.

Mullan (1984) suggested that the empirical saturated limits on $L_X/L_{bol}$ are due to an upper limit on the mechanical energy flux $F_{mech}$ which is generated by convective motions. However, in attempting to evaluate the ratio $F_{mech}/F_{bol}$, the calculation done by Mullan (1984) was subject to several uncertainties. First, the numerical values of the convective speeds $V$ in cool dwarfs were not known reliably. Second, the numerical values of specific heat $C_p$ of the partially ionized gas in a stellar convection zone were greatly overestimated: $C_p$ was assumed to exceed the classical value of $(5/2)\Re/\mu$ by a factor of 27. This erroneous overestimate arose because of reference to an idealized calculation where the actual physical conditions of a convection zone were not incorporated in a realistic manner. Third, in evaluating $F_{mech}$, equal to the product of an energy density 0.5 $\rho V^2$ times a propagation speed $V_p$, it was assumed that the value of $V_p$ was equal to the convective speed $V$. That is, no particular wave mode was assigned to transport the mechanical energy: the latter was simply provided by the ram pressure of the convective motions. Thus, $F_{mech}$ was set equal to 0.5 $\rho V^3$ erg cm$^{-2}$ s$^{-1}$. In the context of the "survivability" condition mentioned in the previous section, there is a problem with the ram pressure assumption: gas ejected from the convection zone upwards at speed $V$ will move ballistically, and will therefore "survive" only up to altitudes of order $h \approx V^2/(2g)$ above the launch point. In the presence of overshooting, the launch point could be as high as a few hundred km above the photosphere. In the cool stars of interest to us here, $g$ is at least as large as the solar value 2.7 x 10$^4$ cm s$^{-2}$. Therefore, even if we assign an optimal value $V$ as large as 1 km s$^{-1}$ (this is a significant overestimate, as we shall see below), we find that $h$ has a value of no more than a few km above the launch point. This is far short of the linear dimensions required (10$^5$ km) to reach a stellar corona. For these reasons, we believe that it is now appropriate to re-visit the conclusions of Mullan (1984) using more realistic models of convective properties.

## 5. MAGNETIC FIELDS GENERATED BY EQUIPARTITION WITH CONVECTIVE MOTIONS

In a medium where the magnetic field strength is $B$ Gauss, and the gas density is $\rho$ g cm$^{-3}$, the flux of energy carried by Alfven waves can be estimated as of order $F_{Alf} = 0.5\ \rho\ (\delta v)^2\ V_A$. Here, $\delta v$ is the r.m.s. turbulent speed which is imposing transverse displacements on the field lines, and $V_A = B/\sqrt{(4\pi\rho)}$ is the Alfven speed. Because of the inherently 3-dimensional nature of convection, we assume that $\delta v$ is essentially the same as the convective speed $V$ which is computed in the model: as convective flows circulate around a cell, the vertical speed ($V$) is comparable to the horizontal speed $\delta v$ at the top and bottom of the cell where the vertical flows spread out (or converge) in preparation for completing the cycle of circulation of matter around the cell. In the presence of a (more or less) vertical field, $\delta v$ serves as the vehicle to impose transverse motion on a field line, thereby setting up the conditions for launching a vertically propagating Alfven wave.

Now we introduce a key assumption of the present calculation: the dynamo in a turbulent convection zone generates magnetic fields $B_{eq}$ which are in equipartition with the convective kinetic energy. Support for this assumption can be found in several papers. (a) Dobler et al. (2006) reported on 3-D MHD models of convective spheres, in which they found that, in the faster rotating runs, the energy in magnetic form grows from starting values of essentially zero to values which, at least near the surface, are comparable to the kinetic energy of the convective motions. In the words of Dobler et al.: "The dynamo-generated magnetic field saturates at equipartition field strength near the surface". (b) Browning (2008) subsequently reported on 3-D MHD models of convection zones in which he finds that fields are generated with strengths close to $B_{eq}$. (c) Yadav et al. (2015) have reported on anelastic MHD models of fully convective M dwarfs: a distributed dynamo generates magnetic fields which lead to a predominantly dipole component at the surface. The authors report that "The magnetic field grows exponentially until its energy is in rough equipartition with the kinetic energy".

In view of these papers, we consider it reasonable to assume in the present work that the field strengths which are generated in convective flows in M dwarfs are given by $B_{eq}^2/8\pi = 0.5\, \rho V^2$. In this case, the Alfven speed (defined by $V_A = \sqrt{B_{eq}^2/4\pi\rho}$ ) reduces to a particularly simple form: $V_A = V$, the convective speed. Combining these results with the definition of $F_{Alf}$ given above, we find that in a cool dwarf, the mechanical flux $F_{mech}$ of energy in the form of Alfven waves is given by $F_{mech} = 0.5\, \rho V^3$ erg cm$^{-2}$ s$^{-1}$.

Interestingly, this is the same as the formula used by Mullan (1984). However, in the present case, there is a significant difference in interpretation. In Mullan (1984), one factor of $V$ in the $V^3$ term represents the upward motion $V$ of a "blob" of gas launched upward from the top of the convection zone. Such a blob behaves ballistically, and suffers serious deceleration by gravity within a height of order $h \approx V^2/(2g)$. Because of the ballistic nature of the motion, the blob does not gain any more energy once it has been launched. By contrast, in the present paper, one of the factors of $V$ is associated with Alfven wave propagation. This factor is by no means subject to ballistic effects. Instead, in the case of an Alfven wave, the wave is subject to restoring forces due to the tension of the local magnetic field along which the wave propagates. As a result, the magnetic field provides a source of energy to help the wave to continue their upward propagation. This source of energy has no analog in the ballistic case. Thus, even though the wave and the blob are both launched with the same speed ($V_A$ and $V$ respectively, where $V_A = V$ in the present case), the "blob" cannot propagate far before it stops, but the wave picks up energy from the field and as a result can propagate a long distance before it dissipates.

6. RESULTS OF MODELING

In order to rectify the errors in the approximate treatment of Mullan (1984), rather than merely guessing what the numerical values of various parameters might be, the approach here is to use detailed models of low-mass stars which are evolved from pre-main-sequence stages to the main sequence. Then a snap-shot of the outer regions of the star is examined when the star reaches the main sequence. We are particularly interested in the radial profiles of 4 physical parameters: $C_p$, $V$, $F_{Alf}$, and $B_{eq}$.

The stellar models are computed using a code which has been developed by one of us (JM): details of the code, including aspects of the choices of equation of state and boundary conditions can be found in Mullan et al. (2015: MHM). In all models we use the equation of state of Saumon, Chabrier & van Horn (1995: SCVH) for hydrogen – helium mixtures augmented with a correction to allow for the presence of heavy elements. We have made 3 sets of models to test the sensitivity of our results to treatment of boundary conditions and choice of mixing length ratio. For set (i) we use the BT Settl atmosphere models of Allard, Homeier & Freytag (2012) to determine the outer boundary condition, specifically pressure and temperature at optical depth $10^3$. The mixing length ratio for the interior is taken to be the same as used in the atmosphere models, i.e. $\alpha = 1.0$. For set (ii) the outer boundary condition is determined from the Eddington approximation at optical depth 0.1. The mixing length ratio is the same as for set (i), $\alpha = 1.0$. Set (iii) differs from set (ii) only in the value of the mixing length ratio, $\alpha = 1.7$. (This value of $\alpha$ is what our code needs in order to fit the parameters of the current Sun best.) No great differences have emerged between models from sets (i), (ii), or (iii) for a given mass. In view of this result, in what follows we present our results for only one set of models, namely set (ii).

For the present work, models were computed for masses between 0.3 and 1.0 $M_\odot$. Although we are interested in this paper mainly in $F_{mech}$ in M dwarfs, we wanted to see if our models would also replicate certain properties of G dwarfs: hence our extension of the mass range to 1.0 $M_\odot$. Our results did indeed show that, as VW report, $F_{mech}$ in G dwarfs rises to larger values than in M dwarfs. However in what follows, we are particularly interested in the range of masses where M dwarfs on the main sequence undergo a transition to complete convection (TTCC): this occurs at masses of 0.3-0.4 $M_\odot$ (see MHM). As a result, we will present mainly results for that range of masses, with only a passing reference to results in more massive stars.

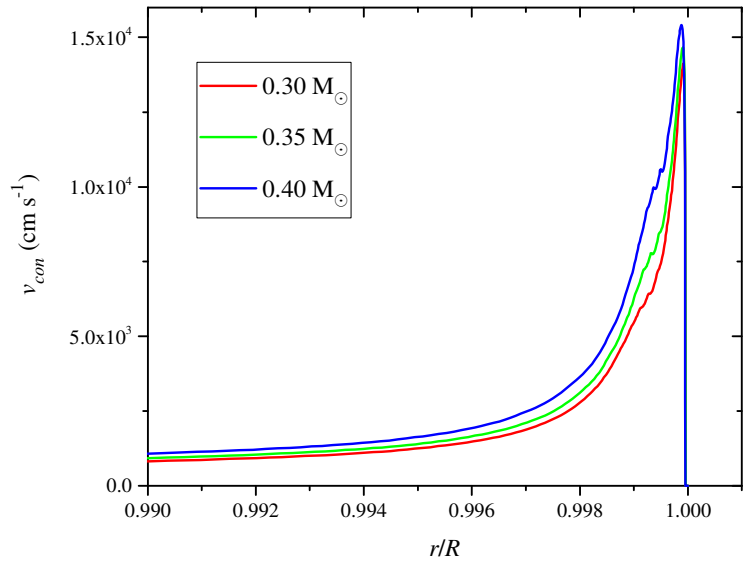

Fig. 1. Radial profiles of convective velocity in the outermost 1% of the radius of models with SCVH(Z) EOS and Eddington boundary conditions with mixing length ratio $\alpha = 1.0$ (type (ii) models). Models have masses of 0.3 - 0.4 $M_\odot$, bracketing the location of the TTCC. Convective velocities do not exceed 0.15 km s$^{-1}$.

In Fig. 1, the value of $V$ is plotted for type (ii) models with masses 0.3 - 0.4 $M_\odot$ as a function of radial location near the surface. Maximum values of $V \sim 0.15$ km s$^{-1}$ are obtained close to the surface (at $r/R = 0.9999$. For completeness, we note that in a model with 1.0 $M_\odot$, the maximum $V$ is found to be 1.1-1.2 km s$^{-1}$: these values are consistent with reported values of microturbulence (probably associated with convective turbulence) in the solar photosphere (e.g. Mullan 2009, p. 53)

In Fig. 2, the radial profiles of $C_p$ are plotted for models belonging to set (ii) in the region where hydrogen ionization is occurring: the lowest values of $C_p$ are (as expected) close to the classical $2.0 \times 10^8$ erg g$^{-1}$ K$^{-1}$, while the largest values exceed the classical value by values of no more than 6 - 8. Thus, the use of an excess by a factor of 27 over and above the classical value assumed by Mullan (1984) was in error by more than a factor of 3 - 4.

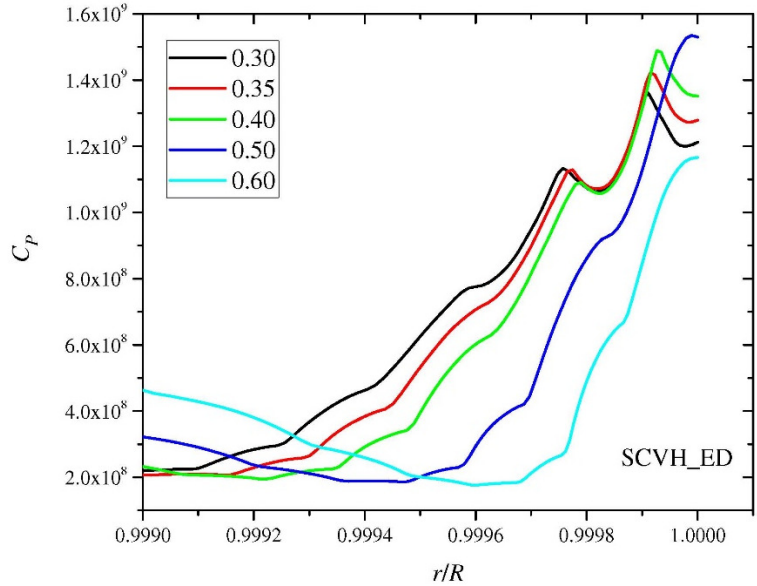

Fig. 2. Radial profiles of the specific heat (in units of erg g$^{-1}$ K$^{-1}$) in the outermost 0.1% of the radius of models with Eddington approximation boundary conditions and mixing length ratio $\alpha = 1.0$. The classical value of $C_p$ in the units used for plotting is $2.0 \times 10^8$.

In Fig. 3, radial profiles of $B_{eq}$ are plotted for models belonging to set (ii). At the radial locations where we shall find $F_{mech}$ rising to maximum values, ce, i.e. at $r/R \approx 0.999$ (see Fig. 4), the $B_{eq}$ values are found to be in the range 250-350 G for stars with masses of 0.3- 0.4 M$_\odot$. Is there any way to decide whether or not these field strengths are plausible for M dwarfs? We refer to independent work which we have done on obtaining global models of M dwarfs in the context of a particular model of magnetoconvection (MacDonald & Mullan 2014: MM14): for 3 M dwarf systems with the most precisely known masses and radii, our models yield surface field strengths in the range 338-505 G. Without any *a priori* attempt to ensure consistency between the present equipartition work and our earlier magnetoconvection modeling, it is encouraging that the values of $B_{eq}$ obtained in the present paper actually turn out to have some overlap with the results of MM14.

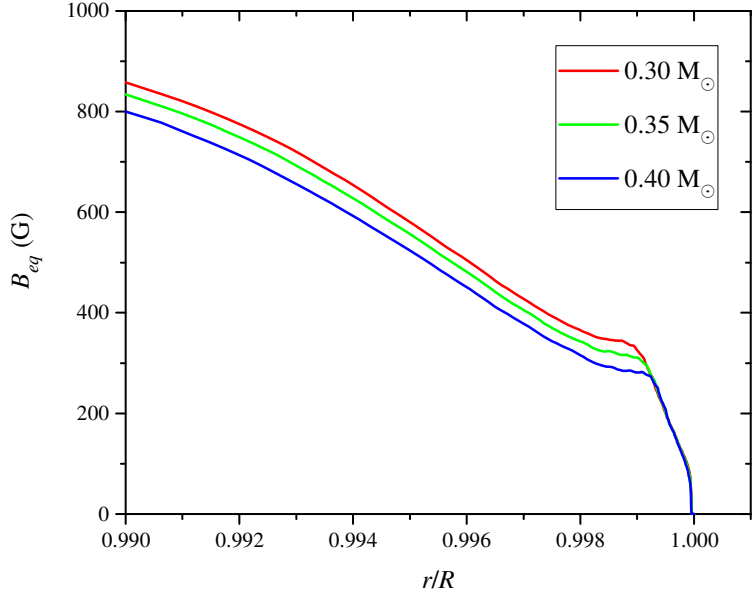

Fig. 3. Radial profile of the equipartition magnetic field strength $B_{eq}$ (in units of Gauss) in the outermost 1% (≈a few thousand km) of the radius of models in set (ii). Models have the same range of masses as in Fig. 1.

Finally, the radial profiles of $F_{mech}$ are plotted for stars of 3 different masses in Fig. 4. The results in the figure refer to models in set (ii). In all cases, a peak in $F_{mech}$ is found at a radial location $r \approx 0.999\,R$. The amplitude of the peak $F_{mech}$ is of particular interest to us here: the numerical value ranges from 2.3 to 2.8 x $10^7$ erg cm$^{-2}$ s$^{-1}$.

Although we do not illustrate results for sets (i) or (iii), those results indicate that, other things being equal, the peak value of $F_{mech}$ scales with the mixing length ratio α roughly as $α^{1.2}$. As a result, for models in set (iii), the peak values of $F_{mech}$ are 1.9 times larger than the range for set (ii). Thus, allowing for uncertainties in the "true" value of α, a realistic range of peak values of $F_{mech}$ predicted by our models of 0.3-0.4 M$_\odot$ stars is (2-5) x $10^7$ erg cm$^{-2}$ s$^{-1}$.

Given stellar radii which (in units of solar radii) are essentially equal to the masses (in terms of a solar mass), the radial location of the peaks in Fig. 4 correspond to linear distances beneath the surface of order 200 - 300 km. At depths of such magnitude in the Sun, Osterbrock (1961) reports that in fields of 50 G, the damping lengths for Alfven waves due to Joule heating are 200 - 400 km. (Frictional dissipation is much weaker, according to Osterbrock, and can be neglected as regards damping of Alfven waves in our case.) To the extent that electron densities are not greatly different in the low-mass stars we study here, Osterbrock's results suggest that a flux $F_{mech}$ of Alfven waves generated at depths of 200 - 300 km in our stars have a good chance of reaching the surface with a flux which is not significantly reduced below $F_{mech}$.

The linear depths of order 200-300 km mentioned in the previous paragraph are based on the assumption that stellar radii (in solar units) are essentially equal to mass (in solar units). This is a good approximation for non-magnetic models of low-mass stars on the main sequence (e.g. Bressan et al. 2012). But in the presence of global magnetic fields, the models of MM14 suggest that a star undergoes some "bloating", and takes on a somewhat larger radius than in a non-magnetic model of the same mass. However, the magnitude of the "bloating" in the MM14 models is in all cases found to be no more than a few percent of the non-magnetic radius. As a result, if we were to attempt to refine our estimates of the linear depths to accommodate the "bloated" magnetoconvection models, our estimates of the linear depths would shift from 200-300 km to perhaps 210-315 km. In the context of Osterbrock's results, these slight changes in linear depth would have no significant effect on our conclusions.

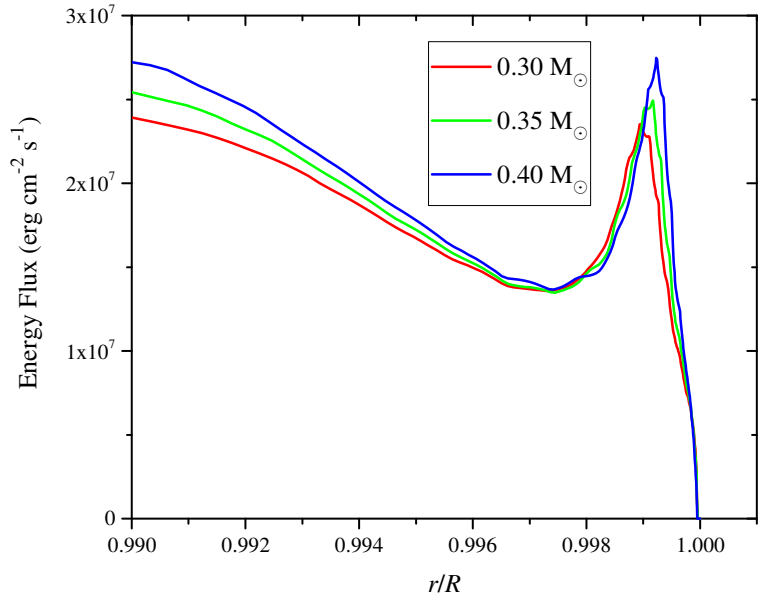

Fig. 4. Radial profiles of Alfven energy flux in the outermost 1% of the radius of 3 models bracketing the TTCC. The models plotted here belong to set (ii), i.e. Eddington approximation, with mixing length parameter $\alpha = 1.0$.

In view of these results, we suggest that the Alfven waves which are generated in a dynamo where magnetic fields are in equipartition with convective flows in low-mass stars of 0.3 - 0.4 $M_\odot$ have access to the corona with mechanical fluxes in the range (2-5) x $10^7$ erg cm$^{-2}$ s$^{-1}$. We note that these predicted fluxes of Alfven waves overlap significantly with the empirical upper limits, (1.4 -3) x $10^7$ erg cm$^{-2}$ s$^{-1}$, reported by VW for coronal heating in the most active M dwarfs within 25 parsecs.

The models of low-mass stars which have been calculated with the code used in the present work indicate that convective motions are energetic enough to (i) generate strong enough fields, (ii) facilitate large enough Alfven speeds, and (iii) generate Alfven waves with copious enough fluxes to serve as a source of the mechanical energy which M dwarfs need to account for "saturated" coronal emission which is observed at the level of 0.2% of the bolometric power.

7. CAN WE DISTINGUISH TURBULENT DYNAMOS FROM INTERFACE DYNAMOS?

A quantitative model for "interface dynamos", i.e. rotationally driven dynamos in stars with radiative cores, has recently been proposed (MHM). The model is based on using observed mean rotational periods to estimate field strengths at the interface between radiative and convective regions. In that work, the assumption was that equipartition exists between magnetic energy and the kinetic energy of *rotation* (rather than convection). For stars that overlap in masses with the ones we study in the present paper (0.44 and 0.33 $M_\odot$), the fields *at the interface* were estimated to be 0.7 - 1.0 MG. Using these values as a starting point, MHM wished to estimate how strong the fields would be when the flux ropes would have risen up to the surface. To do this, they noted that flux ropes rising from the interface have to pass through gas of decreasing density in order to reach the surface. As the ambient density decreases, flux ropes expand, and this causes $B$ to diminish. Cheung & Isobe (2014: CI14) report that $B$ should decrease toward the surface according to a power law of density: $B \sim \rho^k$. Referring to numerical models of solar magnetic ropes, CI14 state that $k = 1.0$ deep inside the Sun, whereas a more appropriate value of $k$ would be 0.5 close to the surface. A flux rope starting at the interface and rising to the surface would (according to CI14) likely undergo a transition from $k = 1$ to $k = 0.5$ along the way. MHM considered these two extreme values for $k$ in order to estimate the surface field strengths corresponding to interface fields of 0.7-1.0 MG. In the case $k = 0.5$, MHM found surface fields $B_s$ = 131-326 G. In the case $k = 1$, $B_s$ = 2 - 6 G.

It is highly likely that the rising flux ropes in a "real star" will not experience a single value of $k$ during the entire journey to the surface. Instead, different values of $k$ will surely occur during the journey to the surface, depending on how much distortion occurs along the way as the flux rope tries to make its way through the convective turbulence. In view of that, it is not realistic to expect that we could draw any strong conclusions from the above cases where $k$ takes on a unique value from interface to surface. Nevertheless, it is noteworthy that the MHM estimates of surface fields (ranging up to 326 G) overlaps well with the "surface" fields we have estimated here for a convective dynamo: 250-350 G (see Figure 3). This overlap suggests that even if two distinct dynamo processes are actually at work in stars, the surface manifestations of the fields might in the "real world" overlap as regards the observational signatures of chromospheric/coronal emission. Thus, even if the dynamo were to undergo a real physical switch in its mechanism from interface to turbulent at the TTCC, there might not be any clear signature of a change in the level of coronal emission. In fact, observations do not reveal any such signature (see e.g. Mullan & MacDonald 2001). An anonymous referee has suggested that, in view of recent 3D MHD simulations of convective shells by Nelson et al. (2013) and Brown et al. (2011), in which "wreaths" of magnetic field are found to be generated by turbulent dynamos in fast rotators (where no interface exists), it is possible that in some stars, an interface dynamo and a turbulent dynamo may both be at work simultaneously.

However, we think it is worthwhile to consider the question: do we have any hope that we might be able to distinguish between interface and turbulent dynamos in cool dwarfs? Durney et al. (1993) have stressed that the relationship between magnetic field and rotation in the presence of an interface dynamo differs from the relationship (if any) between magnetic field and rotation in the presence of a turbulent dynamo. On the one hand: "The solar cycle magnetic field [is] generated at the …interface…needing helicity (or rotation) for its generation". On the other hand: "no rotation, or helicity, is needed for the generation of turbulent field". According to the conclusions of Durney et al., we expect that a pronounced rotation-activity correlation (RAC) should exist in stars where interface dynamos are at work, but the correlations should be less pronounced (or even absent altogether) if turbulent dynamos are at work.

In fact, empirical evidence for a pronounced RAC is well established among FGK dwarfs (e.g. Noyes et al. 1984). These are precisely the stars in which an interface is known to exist, and therefore an interface dynamo *could* be operative. According to Durney et al. therefore, we would not be surprised to find that there exists a pronounced RAC in FGK stars. If it were to be true that turbulent dynamos are actually dominant in completely convective stars (beyond the TTCC), the hypothesis of Durney et al. would suggest that stars beyond the TTCC should exhibit little or no signs of RAC.

If Durney et al. (1993) are correct, then discovery and quantification of *changes (if any) in the RAC's* might help us to distinguish between dynamo modes. In such an event, examination of the detailed structure of RAC's might be the best way forward. The most pertinent feature may be the *slope* of the RAC. If it were possible to sample stellar dynamos belonging to distinct modes, then a slope that is statistically distinct from zero could be consistent with the occurrence of an interface dynamo, whereas a slope of zero would be more consistent with a turbulent dynamo. Unfortunately, the picture is actually more complicated. As VW have shown, dynamos in certain low mass stars are observed to enter into a saturated regime (especially in fast rotators): in such cases, the slope of the RAC tends towards zero (see VW) simply because of saturation. In view of this complication, even if one were to detect a zero slope for the RAC of a sample of stars, this would not necessarily constitute an unambiguous signature of a turbulent dynamo: it might simply be a signature of saturation. In this regard, we note that West et al. (2015) have reported on a search for the RAC slopes by dividing a sample of M dwarfs into two broad groups: M1 - M4 and M5 - M8: they report a slope *a* of -0.016 ± 0.050 (consistent with zero) for the M5-M8 sample. However, they report that almost 90% of the stars in their M5-M8 sample are "active" stars (as indicated by strong H-α emission): such active stars are almost certainly rotating fast, with rotation periods which are short enough to be well into the saturated part of the RAC. This raises the possibility that saturation may be contributing significantly to the zero slope of the M5-M8 sample. For the M1-M4 stars, many (65%) of their sample are also "active". It is true that the RAC slope in the M1-M4 stars is reported as having a finite (negative) slope (*a* = -0.19±0.036), and this might at first sight appear to be proof of a rotationally controlled dynamo (such as an interface dynamo). However, some caution is called for: the slope *a* = -0.19 is much shallower than what has been reported by Wright et al. (2011) for unsaturated stars (*a* = -1 or steeper). Thus, even among the M1-M4 stars, saturation may still be obscuring the true RAC slope.

In order to avoid complications due to saturation, it would be preferable to sample stars in which the dynamo is known to be operating in an unsaturated regime, i.e. in slowly rotating stars. If RAC's could be constructed on either side of the TTCC with a fine-grained choice of spectral sub-types of stars which are known to be *slow* rotators, it might be easier to interpret a change (if any) in the slope of the RAC at the TTCC in terms of a switch from one dynamo mode to another.

Alternatively, it might turn out that the suggestions originally made by Durney et al. (1993) have to be re-visited. In view of the results reported by Brown et al. (2011) and by Nelson et al. (2013), fast rotation of a turbulent convection zone (without any interface) can lead to "wreaths" of magnetic field. In such cases, faster rotation presumably leads to stronger coronal signatures. If this is the case, then an RAC may exist even for this "turbulent dynamo".

8. CONCLUSIONS

We have obtained quantitative estimates of mechanical energy fluxes $F_{mech}$ which emerge in the form of Alfven waves from models of low mass stars in the vicinity of the transition to complete convection. In our model, it is assumed that magnetic fields are generated in equipartition with convective kinetic energy. Our models are found to generate values of $F_{mech}$ which range upto (2-5) x $10^7$ erg cm$^{-2}$ s$^{-1}$. These fluxes overlap well with the upper limits which have been reported for the fluxes required to explain the empirical X-ray emission from the coronae of M dwarfs, (1.4-3) x $10^7$ erg cm$^{-2}$ s$^{-1}$). As a result, we would like to suggest that a turbulent dynamo, in which magnetic fields are generated in equipartition with convective energy, *can* generate enough mechanical flux to account quantitatively for the upper limits on coronal heating in M dwarfs.

In an earlier paper (MHM), we presented numerical evidence that *rotational* energy could account for dynamo activity among stars where an interface exists between an outer convective envelope and an inner radiative core. Such an interface is expected to exist only in stars with masses larger than a critical value (0.32-0.34 $M_{\odot}$). However, in contrast, turbulent dynamos can exist in all stars where convection is operating near the stellar surface. In some stars, both types of dynamos may contribute significantly to the overall mechanical flux. We suggest that in order to distinguish between interface dynamos and turbulent dynamos, the slope of the rotation-activity correlation *might* be a helpful parameter if it can be derived for a sampling of *slowly* rotating ("unsaturated") M dwarfs with closely spaced spectral sub-types. Unfortunately, in practice, the situation may be too complicated to allow a clean distinction.

**Acknowledgements**

DJM is partially supported by NASA DE Space Grant.REFERENCES

Allard, F., Homeier, D., & Freytag, B. 2012, Roy. Soc. London Philos. Trans. Ser. A, 370, 2765

Belcher, J.W. & Davis, Jr., L. 1971, J. Geophys. Res., 76, 3534

Bookbinder, J.A. 1985, PhD thesis, Harvard University.14